\documentclass[aps,prb,twocolumn]{revtex4}
\usepackage{amsmath,amsfonts,graphicx}

\def\ket#1{\left| #1 \right\rangle}

\newcommand{\aver}[1]{\left\langle #1 \right\rangle}

\begin{document}

\title{Many-body interaction in semiconductors probed with 2D Fourier spectroscopy}

\author{Mikhail Erementchouk}
\author{Michael N. Leuenberger}\email{mleuenbe@mail.ucf.edu}
\affiliation{NanoScience Technology Center and Department of Physics, University of Central
Florida, Orlando, FL 32826}

\author{L. J. Sham}
\affiliation{Department of Physics, University of California San
Diego, La Jolla, CA 92093-0319}

\begin{abstract}
A particular difficulty in studying many-body interactions in a solid is
the absence of an experimental technique that can directly probe their key
characteristics. We show that 2D Fourier spectroscopy provides an
efficient tool for the measurement of critical parameters describing the
effect of many-body interactions on the optical response of
semiconductors. We develop the basic microscopic theory of 2D Fourier
spectroscopy of semiconductors in the framework of the three-band model
(heavy holes, light holes, and electrons). The theory includes many-body
correlations nonperturbatively and can be generalized straightforwardly in
order to describe 2D Fourier spectra obtained in atomic physics. We
establish a relation between the 2D Fourier spectrum and the many-body
correlations. It is shown, in particular, that 2D Fourier spectroscopy
provides a principal possibility to establish experimentally the origin of
the fast decay of the memory term describing the Coulomb interaction
between heavy- and light-hole excitons. The theory is applied to an
analysis of the available experimental data. Experiments providing more
detailed information are suggested.
\end{abstract}


\maketitle

Understanding many-body interactions in solids is one of the key problems
of modern solid state physics (see, for instance, the recent review
Ref.~\onlinecite{CHEMLA:2001}). The long-range Coulomb interaction between
electrons leads to complex dynamics of the excitations in semiconductors
and plays the principal role in the nonlinear optical response. Different
experimental techniques have been developed for studying the effects of
many-body interactions. One of the most popular experiments is based on
the four-wave mixing (FWM)\cite{Mukamel}. In these experiments the sample
is illuminated by rays characterized by (non-parallel) wave vectors
$\mathbf{k}_1$, $\mathbf{k}_2$, and $\mathbf{k}_3$. The outgoing signal is
detected in the direction that corresponds to the nonlinear coupling of
the excitation pulses, say, $-\mathbf{k}_1+\mathbf{k}_2+\mathbf{k}_3$. The
advantage of such measurements is that the many-body contribution to the
signal is not blurred by the strong linear (single-particle) component.
The sensitivity of the FWM spectrum to the details of the interaction
between the excitons and other many-body excitations makes it an efficient
tool for probing the many-body properties. However, the standard FWM
experiment does not allow one to make a distinction between different
contributions of many-body interactions and correlations to the shape of
the resonance. As a result it is difficult to interpret a FWM spectrum and
to extract specific characteristics of the many-body interactions and
correlations.

Recently the more flexible technique of two-dimensional Fourier
spectroscopy\cite{Fourier2DGeneral,JonasSpectroscopy,Fourier2DAtomic}
has been applied to studying the semiconductor
properties\cite{Fourier2DOE,Fourier2DPRL}. The general scheme is
similar to that of standard time-resolved FWM experiments with three
pulses propagating along $\mathbf{k}_1$, $\mathbf{k}_2$, and
$\mathbf{k}_3$ launched at $t=t_1$, $t_2$, and $t_3$, respectively.
The difference from a standard FWM experiment is that measurements
are performed not at a fixed or just a few values of the delay time
$\tau = \min(t_2, t_3) - t_1$, but rather for a dense series of
values lying in some interval. Subsequently, the Fourier transforms
are done with respect to the delay time as well as with respect to
the signal time. These two Fourier transforms constitute the
two-dimensional Fourier spectrum. It is important to emphasize that
in the experiments reported in
Refs.~\onlinecite{Fourier2DOE,Fourier2DPRL} the difficult problem of
measuring both the real and the imaginary parts of the signal was
resolved. These experiments, therefore, provide the information
about the phase acquired during the delay time.
As will be shown below, this allows one to make a distinction between the
diffraction on the gratings created by the heavy- and the light-hole
excitons. Thus, 2D Fourier spectroscopy experiments give substantial
insight into the details of the many-body correlations and provide vital
information, which is barely accessible using the standard approach.
Despite many advantages provided by the 2D Fourier spectroscopy, the
application of this technique suffers from the lack of the understanding
of the spectra from the microscopic standpoint. Usually, 2D Fourier
spectra are described in the framework of the phenomenology of the
nonlinear susceptibility, which hides the relation between features of the
spectra and the microscopic characteristics of the system.

Here we develop the basic microscopic theory of 2D Fourier spectroscopy of
the semiconductors in the framework of the three-band model (heavy holes,
light holes, and electrons). We use the perturbational approach with
respect to the excitation field. The interaction between the excitons, on
the contrary, is taken into account exactly. We show that the 2D Fourier
spectrum gives the unique opportunity to measure the key quantities
describing the exciton-exciton interaction. We use our theory to analyze
the experimental results of Ref.~\onlinecite{Fourier2DOE}. Our
calculations produce relations between the spectral features of the 2D
Fourier spectrum and the parameters characterizing the many-body
interaction that allow us to suggest experiments that would provide more
detailed information. Because of the generality of the approach, our
theory is able to describe also the 2D Fourier spectra of other physical
systems, such as molecular nanostructures.

The basic idea behind the derivation of the equations of motion of the
exciton polarization is that the states with the definite number of
particles are the eigenstates of the Hamiltonian of the unperturbed (i.e.
without external field) semiconductor. Initially, before the first pulse
hits the sample, the system is assumed to be in the vacuum state $\ket{0}$
with full valence and empty conduction bands. The excitation pulses couple
the eigenstates of the system populating, thereby, states with different
number of particles. The excitation pulses create directly
$P^{(l)}_\kappa$, the linear response polarization. This excitation is
described in the rotating-wave approximation by
\begin{equation}\label{eq:linear_response_perturbation}
 \left(\frac{\partial}{\partial t} + i\omega_\kappa +
  \Gamma_\kappa\right)P^{(l)}_\kappa = -d
  E^{(l)}_{\sigma_\kappa}(t),
\end{equation}
where $l=1,2,3$ enumerates the exciton pulses according to their time
order. Here and below the Greek letters $\kappa$, $\lambda$, $\mu$, $\nu$
are multi-indices denoting the exciton state according to $\kappa =
\{n_\kappa, \sigma_\kappa\}$, where $n_\kappa$ is the type of the exciton
equal to $\mathrm{h}$ or to $\mathrm{l}$ for heavy-hole and light-hole
excitons, respectively, and $\sigma_\kappa$ is the helicity of the state.
In Eq.~(\ref{eq:linear_response_perturbation}) we have introduced
$\Gamma_\kappa$ as the decay rate of the exciton state $\kappa$, and
$\omega_\kappa$ as the detuning, which is the difference between the
frequency of the rotating frame and the exciton frequency
$\widetilde{\omega}_\kappa$. The external source is specified by the
dipole moment $d$ and by the component of the $l$-th pulse,
$E^{(l)}_{\sigma_\kappa}(t)$, with the helicity $\sigma_\kappa$. The
linear responses $P^{(l)}_\kappa$, in turn, serve as sources of the
third-order polarization. The dynamics of the third-order polarization is
conveniently written in terms of the operator $D_{\mu\nu} = [B_\mu,[B_\nu,
H]]$, where $H$ is the Hamiltonian of the unperturbed semiconductor and
the operator $B^\dagger_\mu$ creates an exciton in the state $\mu$. The
dynamics of the polarization corresponding to the FWM signal in the
direction $-\mathbf{k}_l + \mathbf{k}_m + \mathbf{k}_n$, where $l$, $m$,
and $n$ enumerate the excitation pulses, is governed
by\cite{OSTREICH:1998}
\begin{widetext}
\begin{equation}\label{eq:FWM_perturbation_total}
\begin{split}
  \left(\frac{\partial}{\partial t} + i\omega_\kappa +
  \Gamma_\kappa\right)P_\kappa = & \sum_{\lambda,\mu,\nu}\left\lbrace
  -i \beta^{\kappa\lambda}_{\mu\nu} {P^{(l)}_\lambda}^* P_\mu^{(m)}
  P_\nu^{(n)}\right.
  + \frac{1}{2}
  {P^{(l)}_\lambda}^*\int_{t_0}^t dt'e^{-(\Gamma_\mu+\Gamma_\nu)(t-t')}
  F^{\kappa\lambda}_{\mu\nu}(t-t')P_\mu^{(m)}(t') P_\nu^{(n)}(t') \\
 & \left. - i d C^{\kappa\lambda}_{\mu\nu} {P^{(l)}_\lambda}^*
 \left[P_\mu^{(m)}E_{\sigma_\nu}^{(n)}(t)+ P_\mu^{(n)}E_{\sigma_\nu}^{(m)}(t)\right]
  \right\rbrace.
\end{split}
\end{equation}
\end{widetext}
The parameters $\beta$ and the memory functions $F$ describe the
effect of the exciton-exciton correlations. Obviously, interactions
involving higher number of particles do not contribute to the
third-order polarization. These parameters are defined by
$\beta^{\kappa\lambda}_{\mu\nu} =
\aver{D_{\kappa\lambda}B^\dagger_\mu B^\dagger_\nu}$ and
$F^{\kappa\lambda}_{\mu\nu}(\tau) =
\aver{e^{iH\tau}D_{\kappa\lambda}e^{-iH\tau}D^\dagger_{\mu\nu}}$.
The last term in Eq.~(\ref{eq:FWM_perturbation_total}) accounts for
the Pauli blocking with the phase-space filling parameters\cite{OSTREICH:1998}
$C^{\kappa\lambda}_{\mu\nu}$.

As follows from Eqs.~(\ref{eq:linear_response_perturbation}) and
(\ref{eq:FWM_perturbation_total}) we need to take into account only
the excitons with helicity of unit magnitude, i.e. the ones coupled
to the one-photon states of electromagnetic field. Therefore, in
what follows, we will consider only excitons with
$\sigma_\kappa=\pm1$.

We would like to emphasize that Eq.~(\ref{eq:FWM_perturbation_total})
exactly accounts for the Coulomb interaction between excitons. The
generality of the basic ideas and non-restrictive assumptions make this
equation virtually model-independent. However, in order to present the
relation between the 2D Fourier spectrum and the parameters characterizing
the many-body interaction in the most transparent way, we restrict
ourselves to the short-memory approximation.\cite{OSTREICH:1998} The
effect of non-locality of the memory function on the 2D Fourier spectrum
will be investigated elsewhere. The short-memory approximation corresponds
to relatively short biexciton life time\cite{OSTREICH:1998} or fast decay
of the memory function.\cite{SAVASTA:2003_PRL,SAVASTA:2003_SST} In the
framework of this approximation the last term in the r.h.s. of
Eq.~(\ref{eq:FWM_perturbation_total}) is substituted by an instantaneous
term, which we absorb into the modified $\beta$-parameter defining
\begin{equation}\label{eq:modified_beta}
  \widetilde{\beta}^{\kappa\lambda}_{\mu\nu} = \beta^{\kappa\lambda}_{\mu\nu} +
  \gamma^{\kappa\lambda}_{\mu\nu}.
\end{equation}
In what follows we will refer to $\widetilde{\beta}$ as the modified
mean-field parameter, outlining that it is local in time.

The exact form of $\gamma^{\kappa\lambda}_{\mu\nu}$ depends on the
origin of the fast decay of the kernel. In the simplest case of
short life-time one has\cite{OSTREICH:1998}
\begin{equation}\label{eq:gamma_short_life}
 \gamma^{\kappa\lambda}_{\mu\nu} = \frac{i}{2(\Gamma_\mu +
 \Gamma_\nu)} F^{\kappa\lambda}_{\mu\nu}(0).
\end{equation}
In the case of the fast decay of the memory
function\cite{SAVASTA:2003_PRL,SAVASTA:2003_SST} the modification of
the $\beta$-parameter takes the form
\begin{equation}\label{eq:gamma_fast_decay}
 \gamma^{\kappa\lambda}_{\mu\nu} = \frac{1}{2}
\aver{D_{\kappa\lambda}\frac{1}{H - i(\Gamma_\mu +
 \Gamma_\nu)}D^\dagger_{\mu\nu}}.
\end{equation}

We study the polarization mixing induced by the many-body
interaction using the standard three band (electrons, heavy, and
light holes) semiconductor model, where the exciton destruction
operator has the form $B_\mu = \int dx dy\,B_\mu(x,y)$ with
\begin{equation}\label{eq:B_mu}
  B_\mu(x,y) = \phi^*_\mu(x - y) v^\dagger_{\sigma_\mu(1)}(y)
  c_{\sigma_\mu(2)}(x).
\end{equation}
Here $\phi_\mu(x - y)$ is the exciton envelope wave function, $v$ and $c$
are the annihilation operators acting on the states in the valence band
and the conduction band, respectively. The indices $\sigma_\kappa(i)$
specify the spin of the electron in the valence ($i=1$) and conduction
($i=2$) bands. In terms of $B_\mu(x,y)$ one can explicitly write
\cite{Sham_JL}
\begin{equation}\label{eq:force_operator}
  D_{\mu\nu}=\int dx_1\ldots dy_2 B_\mu(x_1, y_1) B_\nu(x_2, y_2)
 U(x_1, y_1; x_2, y_2),
\end{equation}
where $U$ is the energy of the electrostatic interaction between two
excitons.

From the definitions of $\widetilde{\beta}^{\kappa\lambda}_{\mu\nu}$
and $\gamma^{\kappa\lambda}_{\mu\nu}$ one can derive the spin
selection rules. It can be shown that the contribution of the
$\beta$-term reduces to $\propto -i \beta_\kappa^{\kappa}
  |P_{{\kappa}}|^2P_\kappa -i \beta_\kappa^{\bar{\kappa}}
  |P_{\bar{\kappa}}|^2P_\kappa$,
where $\bar{\kappa}$ denotes the exciton state ``conjugate" to
$\kappa$, with the conjugation understood according to the rule
$\{\mathrm{h},\pm1\} \leftrightarrow \{\mathrm{l},\mp1\}$. The
$\gamma$-term turns out to be less restrictive having the form
$\propto\sum_{\lambda}\gamma_\kappa^\lambda  |P_\lambda|^2P_\kappa$.
For both $\beta$ and $\gamma$ the reduction of the indices is
performed according to the same rule
$\widetilde{\beta}_\kappa^\lambda =
\widetilde{\beta}_{\kappa\lambda}^{\kappa\lambda}(2 -
\delta_{\kappa\lambda})$. The expressions for $\beta$ and $\gamma$
are combined together to the modified mean-field term according to
Eq.~(\ref{eq:modified_beta}). In what follows we treat
$\widetilde{\beta}_\kappa^\lambda$ as phenomenological parameters.
It suffices for our purposes since we are interested in a relation
between the 2D Fourier spectral features and the microscopic
characteristics rather than in first-principle calculations of 2D
Fourier spectra.


We consider the 2D Fourier spectrum obtained in the so-called
rephasing scheme\cite{BOYD:2002}, when the excitation pulse
corresponding to the conjugated field arrives first, i.e. $l=1$ in
Eqs.~(\ref{eq:linear_response_perturbation}) and
(\ref{eq:FWM_perturbation_total}). Resolving these equations with
respect to the FWM polarization $P_\kappa$ and performing the
Fourier transform with respect to both the signal time $t$ and the
delay time $\tau = \min(t_2, t_3) - t_1$, we obtain for
$P_\kappa(\omega,\Omega)$ the expression
\begin{widetext}
\begin{equation}\label{eq:solution_perturbation_FWM_Fourier}
 P_\kappa =  
 d|d|^2
 \sum_{\lambda}
 \widetilde{\beta}_\kappa^\lambda  \frac{
 {E_{\sigma_\lambda}^{(1)}}^*
 \left[E_{\sigma_\lambda}^{(2)}E_{\sigma_\kappa}^{(3)}e^{-2\Gamma_\lambda T}
 +E_{\sigma_\lambda}^{(3)} E_{\sigma_\kappa}^{(2)}f_\kappa^\lambda(T)\right]
 g_\lambda(\tau_{\max})}{(\omega - \omega_\kappa + i\Gamma_\kappa)(\omega - \omega_\kappa + i\Gamma_\kappa+
 2i\Gamma_\lambda)
 (\Omega + \omega_\lambda + i\Gamma_\lambda)}
 + \Pi_\kappa(\omega,\Omega),
\end{equation}
\end{widetext}
where $\Pi_\kappa(\omega,\Omega)$ is the Pauli blocking contribution
to the 2D spectrum
\begin{equation}\label{eq:Pi_kappa}
  \Pi_\kappa(\omega,\Omega) =  -|d|^2 d \sum_{\lambda, \mu, \nu}
  C^{\kappa, \lambda}_{\mu,\nu}
  \frac{{E_{\sigma_\lambda}^{(1)}}^* E_{\sigma_\mu}^{(3)}E_{\sigma_\nu}^{(2)}
  f_\nu^\lambda(T)g_\lambda(\tau_{\max})}{
  (\omega - \omega_\kappa + i\Gamma_\kappa)(\Omega + \omega_\lambda + i\Gamma_\lambda)}.
\end{equation}
In Equations~(\ref{eq:solution_perturbation_FWM_Fourier}) and
(\ref{eq:Pi_kappa}) the function $f_\kappa^\lambda(T) =
e^{-T[i(\omega_\kappa - \omega_\lambda) + \Gamma_\kappa +
\Gamma_\lambda]}$ describes the dependence of the spectrum on the time
separation of the second and the third pulses $T = t_3 - t_2$, the
frequencies $\omega$ and $\Omega$ correspond to the signal time and the
delay time, respectively, and $E^{(m)}_\sigma = \int dt
E^{(m)}_\sigma(t)$. Deriving
Eqs.~(\ref{eq:solution_perturbation_FWM_Fourier}) and (\ref{eq:Pi_kappa})
we have used the assumption that the pulses are short compared to the
characteristic dynamical time scales determined by $\omega_\kappa$ and
$\Gamma_\kappa$. The function $g_\lambda(\tau_{\max}) = 1-
e^{i\tau_{max}(\Omega + \omega_\lambda + i\Gamma_\lambda)}$, with
$\tau_{max}$ being the maximal reached value of the delay time, accounts
for the finite range of the delay time used in the experiments and
explains the wavy character of the spectrum along the vertical axis.

As follows from Eqs.~(\ref{eq:solution_perturbation_FWM_Fourier}) and
(\ref{eq:Pi_kappa}) the spectrum has resonances in the 2D $(\omega,
\Omega)$-plane at points with the coordinates $(\omega_\mu, -\omega_\nu)$.
It is seen that a particular exciton state $\kappa$ produces only one
resonance along the $\omega$-axis. The resonances along the $\Omega$-axis
are produced by the nonlinear coupling of the exciton state $\kappa$ with
different exciton states. Such a separation between the interaction with
heavy- and light-hole excitons is possible solely due to the structure of
the 2D Fourier spectrum.

Simple comparison of Eqs.~(\ref{eq:Pi_kappa}) and
(\ref{eq:solution_perturbation_FWM_Fourier}) shows that the 2D spectra of
the FWM signal created by the Pauli blocking and the Coulomb interaction
between the excitons have qualitatively different form. The resonances on
the spectrum produced by the Pauli blocking have the Lorentz form along
the vertical and horizontal axes. The reason is that the dependence of the
signal on the signal and delay time is essentially the free evolution of
the polarization created by a short pulse. This evolution has the form of
oscillations, which produce a simple pole after the Fourier transform. At
the same time the resonances created by the Coulomb interaction fall off
asymptotically as $\propto 1/\Omega$ and as $\propto 1/\omega^2$, along
the $\Omega$- and $\omega$-axes, respectively. This is the direct
consequence of the fact that the FWM polarization is continuously excited
by the polarizations of the linear response. The dependence on the signal
time is found as a convolution of the respective Green function with the
source. After the Fourier transform with respect to signal time it yields
the product of Fourier images of the Green function and the source. This
results in the asymptotic form $\propto 1/\omega^2$ because of the
harmonic time dependence of these functions.

More detailed consideration of the spectrum depends on the specific
experimental situation. Keeping in mind the analysis of the experiments
reported in Refs.~\onlinecite{Fourier2DPRL,Fourier2DOE}, we make several
simplifying assumptions. First of all we note that the experimental
spectra corresponding to the rephasing scheme have clear elongation along
the vertical axis. According to the discussion above this means that the
main contribution to the FWM spectrum in these experiments comes from the
Coulomb interaction. We employ this observation by neglecting the Pauli
blocking term $\Pi_\kappa(\omega, \Omega)$ in
Eq.~(\ref{eq:solution_perturbation_FWM_Fourier}). Additionally, we assume
that the helicities of the excitation pulses and the detected signal are
not resolved. Thus, the signal is obtained a sum of contributions
(\ref{eq:solution_perturbation_FWM_Fourier}). Also, we assume that the
basic exciton characteristics do not depend on helicity, so that we have
only two sets of parameters corresponding to light-hole and heavy-hole
excitons. As a result, one ends up with four resonances in the $(\omega,
\Omega)$-plane plane, situated near the vertices of a square. Finally, we
take $T=0$.

For a qualitative analysis we consider the situation when all resonances
contribute to the 2D Fourier spectrum independently, so that near the
point with coordinates $(\omega_\mu, -\omega_\nu)$ we need to keep only
the respective resonances in
Eq.~(\ref{eq:solution_perturbation_FWM_Fourier}). We denote this resonant
contribution by $\mathcal{E}_{\mu}^\nu$. As follows from
Eq.~(\ref{eq:solution_perturbation_FWM_Fourier}) and the assumption about
non-resolved signal helicity $\mathcal{E}_{\mu}^\nu$ is obtained by
summation over different helicities of the excitons with the frequencies
$\omega_\mu$ and $\omega_\nu$. Employing the assumption of independence of
the material parameters on helicities, we obtain
\begin{widetext}
\begin{equation}\label{eq:signal_apex_mu_nu}
 \mathcal{E}_{\mu}^\nu(\omega,\Omega) = i\frac{A_{\mu}^{\nu}}{(\omega - \omega_\mu+ i\Gamma_\mu)
 (\omega - \omega_\mu + i\Gamma_\mu + 2i\Gamma_\nu)(\Omega + \omega_\nu +
 i\Gamma_\nu)},
\end{equation}
\end{widetext}
where
\begin{equation}\label{eq:C_mu_nu_def}
 A_{\mu}^{\nu} = \frac{2\pi l}{nc}\bar{\omega}_{n_\mu}
 d|d|^2 \sum_{\sigma_\mu,\sigma_\nu}
 \widetilde{\beta}_{n_\mu,\sigma_\mu}^{n_\nu,\sigma_\nu}{E^{(1)}_{\sigma_\nu}}^*
 \left(E_{\sigma_\nu}^{(2)}E_{\sigma_\mu}^{(3)} +
 E_{\sigma_\nu}^{(3)}E_{\sigma_\mu}^{(2)}\right).
\end{equation}
We note that as follows from this expression $A_\mu^\nu =
A_{n_\mu}^{n_\nu}$, i.e. these parameters depend only on the type of the
excitons because of the summation over helicities. However, in order to
improve readability of formulas, we keep the general notations where it is
necessary. In Eq.~(\ref{eq:C_mu_nu_def}) $l$ is the thickness of the
sample, $n$ is the refractive index of the material, and $c$ is the speed
of light. The magnitude of $\mathcal{E}_{\mu}^\nu(\omega,\Omega)$ has the
simplest form. It is a product of pole functions. The real and imaginary
parts of the signal demonstrate a more complex structure. For example, the
real part of the resonance, $R_\mu^\nu$, near the point $(\omega_\mu,
-\omega_\nu)$ changes its sign along the curve $R_\mu^\nu(\omega, \Omega)
= 0$, which in a vicinity of $(\omega_\mu, -\omega_\nu)$ can be
approximated by a straight line
\begin{equation}\label{eq:sign_functions}
  R_\mu^\nu(\omega,\Omega) = \arg\left(A_{\mu}^{\nu}\right) +
 \frac{2(\Gamma_\mu + \Gamma_\nu)}{\Gamma_\mu(\Gamma_\mu + 2\Gamma_\nu)}
  (\omega - \omega_\mu)
  + \frac{1}{\Gamma_\nu}(\Omega + \omega_\nu).
\end{equation}

We apply the developed description of the 2D Fourier spectrum for the
analysis of the experimental data obtained in
Refs.~\onlinecite{Fourier2DOE}.
We would like to start from noting that, as follows from
Eq.~(\ref{eq:sign_functions}), the slope of the zero line
$R_{\mathrm{h}}^{\mathrm{h}}(\omega,\Omega) = 0$ is the constant
$-4/3\approx -1.33$. The experimental value of the slope is found to be
equal to $-1.3$. We would like to stress the universality of the slope and
to emphasize that such a good agreement with the experiment is provided by
the fact that the $(\omega_{\mathrm{h}}, -\omega_{\mathrm{h}})$-resonance
is the strongest one and the slope of the zero line is weakly affected by
other resonances. The slope, however, strongly depends on the dynamical
model describing the exciton polarization. In particular, if one takes
into account the Pauli blocking then the slope would take value depending
on the relation between the Coulomb interaction and the Pauli blocking
contributions into the FWM spectrum. The slope takes the value $-4/3$ in
the limit when the Pauli blocking can be neglected.

\begin{table}
\caption{Fitted values of the parameters of the system studied in
Ref.~\onlinecite{Fourier2DOE}. The values of the exciton frequencies
are provided in the stationary frame.}
\begin{tabular}{|c|c|c|c|c|}
  \hline
   $n$ & $\omega_n$, meV & $\Gamma_n$, meV & $A^{\mathrm{h}}_n$ & $A^{\mathrm{l}}_n$\\ \hline
  $\mathrm{h}$ & $1540$ & $1.3$ & $8.1 - i4.6$ & $6.0 + i9.3$ \\
  $\mathrm{l}$ & $1544$ & $1.7$ & $0.7 - i9.9$ & $12.3 + i1.5$ \\
  \hline
\end{tabular}\label{tab:fit_results}
\end{table}

\begin{figure}
  \includegraphics[width=3.75in]{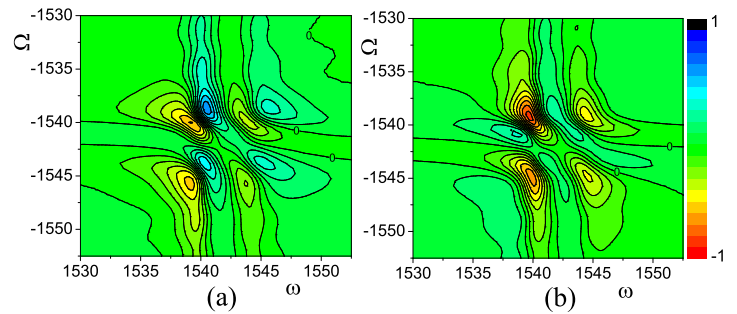}
  \caption{The real part (a) and the imaginary part (b) of the 2D Fourier spectrum
  obtained using the parameters from Table ~\ref{tab:fit_results}.
  These figures
  should be compared with Figs. 7e and 7f of Ref.~\onlinecite{Fourier2DOE}.}\label{fig:spectrum}
\end{figure}

%
%
%




From the experimental data of Ref.~\onlinecite{Fourier2DOE} we find the
parameters $A_{\mu}^{\nu}$ using the least-square method. The exciton
parameters $\omega_\mu$ and $\Gamma_\mu$ are tuned in order to minimize
the deviation between the theory and the experiment. The results of the
fit are shown in Table~\ref{tab:fit_results}. The 2D Fourier spectrum
corresponding to the fitted parameters is depicted in
Fig.~\ref{fig:spectrum}.

The difficulty of extracting more detailed information from the results of
Ref.~\onlinecite{Fourier2DOE} is that the measurements were performed
using linearly polarized pulses and the linearly polarized signal was
detected. Much more detailed information can be obtained if the signals
with circular polarization, i.e. with fixed helicity, are used. As follows
from Eq.~(\ref{eq:C_mu_nu_def}) a particular choice of the helicities of
the excitation pulses and the detected signal allows a direct access to
particular $\widetilde{\beta}$'s. For example, if the excitation pulses
have helicity $\sigma = +1$ and a signal with $\sigma = +1$ is detected,
then each resonance in $(\omega,\Omega)$-plane will be determined by the
specific $\widetilde{\beta}^{n_\nu,+1}_{n_\mu,+1}$. It is interesting to
note that this provides the principal possibility to obtain experimentally
the information about the origin of the fast decay of the memory term
describing the Coulomb interaction between the heavy- and light-hole
excitons.
Indeed, this particular choice of the helicities excludes the contribution
from the interaction between $\{\mathrm{h},+1\}$ and $\{\mathrm{l},-1\}$
excitons, so that only the term $\propto
\gamma_{\mathrm{h},+1}^{\mathrm{l},+1}$ will mostly contribute. If the
memory decay is caused by the short biexciton life-time then, as follows
from Eq.~(\ref{eq:gamma_short_life}), this should result in imaginary
$\widetilde{\beta}^{\mathrm{h},+1}_{\mathrm{l},+1}$. This will not be the
case if the respective matrix elements of $F^{\kappa\lambda}_{\mu\nu}(t)$
decay faster\cite{SAVASTA:2003_PRL} than $1/(\Gamma_\mu + \Gamma_\nu)$.

We would like to note that even more flexible access to different matrix
elements of $\widetilde{\beta}^\nu_\mu$ is provided by the observational
scheme with the time separating second and third pulses equal to the delay
time, $T=\tau$. In this scheme the different sequences of excitation
pulses, which enter symmetrically
Eq.~(\ref{eq:solution_perturbation_FWM_Fourier}), turn out to produce
different resonances along $\Omega$-axis. As a result, the 2D Fourier
spectrum obtained using this scheme has four resonances along the vertical
axis separated by $\omega_{\mathrm{l}} - \omega_{\mathrm{h}}$.

We conclude by considering briefly the effect of inhomogeneous
broadening. The broadening is taken into account by averaging the
spectrum with respect to a joint distribution of the exciton
frequencies. First, we note the different effect of the
inhomogeneous broadening on the diagonal and off-diagonal
resonances. While averaging is performed the resonances situated
near $(\omega_\mu, -\omega_\mu)$ move along the diagonal resulting
in elongating resonances in this direction. At the same time the
width of the resonance in the direction perpendicular to the
diagonal does not change and is determined by the homogeneous
linewidth. As follows from Eq.~(\ref{eq:signal_apex_mu_nu}), the
half-width in the perpendicular direction $\delta_\mu$ of the
magnitude of the signal near $(\omega_\mu, -\omega_\mu)$ is found
from the equation $(\delta_\mu^2 + 2\Gamma_\mu^2)^2 (\delta_\mu^2 +
18\Gamma_\mu^2) = 144\Gamma_\mu^6$ and is equal to $\delta_\mu
\approx 0.88 \Gamma_\mu$. Applying this relation to the spectra
obtained in Ref.~\onlinecite{Fourier2DOE} we obtain $\Gamma_h
\approx 1.2 \mathrm{meV}$ and $\Gamma_l \approx 1.4 \mathrm{meV}$,
which are in a agreement with the values found in
Table~\ref{tab:fit_results}.

The effect of the inhomogeneous broadening on the off-diagonal resonances
is determined by the relation between the distributions of the frequencies
of the heavy-hole and light-hole excitons. For not too high values of the
broadening the value of the light-hole -- heavy-hole splitting can be
considered to be fixed. As a result, averaging leads to elongation of the
off-diagonal resonance along respective line, whose slope depends on the
ratio between the values of inhomogeneous broadenings of the light-hole
and heavy-hole excitons.


In summary, we have developed the basic microscopic theory of 2D Fourier
spectroscopy of semiconductors. We have shown that the resonant
peculiarities in 2D Fourier spectrum are directly related to respective
microscopic quantities describing the exciton-exciton interaction. Because
of the two-dimensional structure of the spectrum the contributions from
the Coulomb interaction between different excitons (heavy- and light-hole)
turn out to be naturally separated. We demonstrate, in that way, that 2D
Fourier spectroscopy provides a unique opportunity to extract the
information regarding many-body correlations in semiconductors from direct
measurements. In particular, we show that it is possible to obtain
experimentally the information regarding the origin of the fast decay of
the memory term describing the Coulomb interaction between the heavy-hole
and light-hole excitons. We have given a simple application of the theory
analyzing the experimental data reported in Ref.~\onlinecite{Fourier2DOE}.

We would like to thank Xiaoqin Li and Steve Cundiff for useful discussions.
This work is supported by NSF DMR 0403465.


\end{document}